\documentclass[bibyear,letter]{aa_old}
\usepackage{graphicx}
\usepackage[varg]{txfonts}
\usepackage{color}

\begin{document}

\title{
  Spatially resolving the AGB star V3 
  in the metal-poor globular cluster 47 Tuc with VLTI/GRAVITY
\thanks{
  Based on observations made with ESO's VLTI/GRAVITY. 
  Program ID: 112.25EN.001
}
}

\author{K.~Ohnaka\inst{1} 
\and
G.~Weigelt\inst{2} 
\and
K.-H.~Hofmann\inst{2} 
\and
D.~Schertl\inst{2}
}

\offprints{K.~Ohnaka}

\institute{
  Instituto de Astrof\'isica, 
  Departamento de Ciencias F\'isicas, 
  Facultad de Ciencias Exactas,
  Universidad Andr\'es Bello, 
Fern\'andez Concha 700, Las Condes, Santiago, Chile\\
\email{k1.ohnaka@gmail.com}
\and
Max-Planck-Institut f\"{u}r Radioastronomie, 
Auf dem H\"{u}gel 69, 53121 Bonn, Germany
}

\date{Received / Accepted }

\abstract
{
Mass loss at the asymptotic giant branch (AGB) plays an important role not
only in the final fates of stars, but also in the chemical evolution of 
galaxies. Nevertheless, the metallicity effects on AGB mass loss are not
yet fully understood.
}
{
  We present spatially resolved observations of an AGB star, V3, 
  in the metal-poor globular cluster 47~Tuc (NGC 104). 
}
{
  The AGB star 47~Tuc V3 was observed using the GRAVITY
  instrument at ESO's Very Large Telescope Interferometer (VLTI) at
  2--2.45~\mbox{$\mu$m,}\ with a projected baseline length of up to 96~m. 
}
{
    The object 47~Tuc V3 has been spatially resolved and stands as  
    the first  to attempt to spatially resolve an
  individual star in a globular cluster. 
    The uniform-disk fit to the observed data results in an
    angular diameter of $\sim$0.7~mas. 
    Our modeling of the spectral energy distribution and
    near-infrared interferometric GRAVITY data suggests that 
    the observed data can be explained by an optically
  thin dust shell with a 0.55~\mbox{$\mu$m}\ optical depth of 0.05--0.25,
  consisting of metallic iron grains, likely together with effects of 
  the extended atmosphere of the central star.
  The dust temperature at the inner shell boundary is 500--800~K
  (corresponding to 23--90 stellar radii), significantly lower than 
  observed in nearby oxygen-rich AGB stars.
  Radiation pressure on small (\mbox{$<$}0.05~\mbox{$\mu$m}) iron grains is
  not sufficient to drive stellar winds. Therefore, iron grains may grow
  to larger sizes, even in the metal-poor environment. Alternatively, it is 
  possible that the observed iron grain formation is a result of the mass outflow
  initiated by some other mechanism(s).
}
{
  The sensitivity and angular resolution of VLTI provides a
  new window onto spatially resolving 
  individual stars in metal-poor globular clusters. This allows us to
  improve subsequent studies of the metallicity dependence of
    dust formation and
  mass loss. 
}


\titlerunning{Spatially resolving the circumstellar envelope of 
the AGB star 47~Tuc V3}
\authorrunning{Ohnaka et al.}
\maketitle

\section{Introduction}
\label{sect_intro}

Low- and intermediate-mass stars experience significant mass loss at the
asymptotic giant branch (AGB), which plays an important role not only
in stellar evolution, but also in the chemical evolution of galaxies. 
To incorporate the mass loss in the stellar evolution theory, we need a
mass-loss formula, which expresses
the mass-loss rate as a function of basic stellar parameters, such as 
the stellar mass, luminosity, temperature, and metallicity. 
However, the observationally derived mass-loss formulae show noticeable
differences due to our incomplete understanding of the mass-loss mechanism
(e.g., Vassiliadis \& Wood \cite{wood93};
van Loon et al. \cite{vanloon05}; Goldman et al. \cite{goldman17}).

To improve our understanding of the metallicity dependence
of mass loss, 
access to observations of stars with different metallicities are crucial.
In this context, AGB stars in metal-poor globular clusters in the Milky Way provide us
with excellent opportunities to work toward this goal. 
Furthermore, because the distance to globular clusters is often well 
known, the luminosity of each star can be determined. 
47~Tuc (NGC~104) is one of the best studied globular clusters at
a distance of 4.5~kpc (Harris
\cite{harris96}\footnote{https://physics.mcmaster.ca/\~{}harris/mwgc.dat}).
Its low metallicity of [Fe/H] = $-0.72$ (Harris \cite{harris96}) is
$\sim$1/5 the
solar value, making it ideal for studying the AGB mass loss at low
metallicities. 
McDonald et al. (\cite{mcdonald19}) analyzed the radio CO observations of 
the pulsating AGB star V3 in 47~Tuc and
found that the terminal velocity is slower
than its Galactic counterparts. This study suggested that the lower metallicity results
in smaller grains, which, in turn, slow down the stellar wind. 

Dust is expected to form at lower temperatures in more metal-poor
environments. 
The modeling of the observed spectral energy distributions (SEDs) of
dusty AGB stars in the metal-poor globular clusters 47~Tuc and 
NGC 362 ([Fe/H] = $-1.16$) 
shows that the dust condensation temperature is 600--1100~K 
(McDonald et al. \cite{mcdonald11}, hereafter Mc11a; 
Boyer et al. \cite{boyer09}). 
These temperatures are much lower than the dust condensation temperature
of $\sim$1500~K observed in the Galactic oxygen-rich AGB stars. 
However, there are ambiguities in the SED fitting, owing
to the degeneracy coming from  many parameters such as the dust condensation
radius, density distribution, grain species, and grain size. 
A straightforward approach to mitigate this degeneracy and better 
constrain the metallicity effects is to
spatially resolve the circumstellar dust envelope.

\section{Observations and data reduction}
\label{sect_obs}

The star V3 is one of the brightest AGB stars in the near-infrared
in 47~Tuc. Its variability with a period of 192 days and
amplitude of $\Delta V \approx 5$~mag, along with the line doubling that is
due to stellar pulsation, indicate that it is a Mira-like variable
(Lebzelter et al. \cite{lebzelter05}). 
We observed 47~Tuc V3 with GRAVITY (GRAVITY Collaboration \cite{gravity17}) at
ESO's Very Large Telescope Interferometer (VLTI) on 2023 October 14
(UTC) at 2--2.45~\mbox{$\mu$m}, using the Auxiliary Telescope (AT) configuration 
D0-G2-K0-J3, with a maximum projected baseline length of $\sim$96~m
(Program ID: 112.25EN.001, P.I.: K.~Ohnaka). 
HD3689 (F6V, uniform-disk diameter = 0.236~mas, 
JMMC catalog: Bourges et al. \cite{bourges17}) and HR9106
(F5V, uniform-disk diameter = 0.226~mas) were observed 
for the interferometric and spectroscopic calibration. 
However, the GRAVITY data of HR9106 show that it is spatially resolved,
although it is expected to be a point source, according to 
its angular diameter from the JMMC catalog. 
Therefore, we only used HD3689 for the calibration of 47~Tuc V3. 
Our GRAVITY observations are summarized in Table~\ref{obslog}.

The GRAVITY data were reduced with the GRAVITY pipeline
ver~1.6.0\footnote{https://www.eso.org/sci/software/pipelines/gravity/}.
The data were originally taken with a spectral resolution of 500. 
To increase the signal-to-noise ratio (S/N)  of
the results, we spectrally binned the raw GRAVITY data (both the
science target and the calibrator as well as the raw calibration
files needed to create the P2VM) with a running box car filter, which
resulted in a spectral resolution of 200.
Each data set of the science target or calibrator consists of two
exposures on the target and one on the sky. We reduced each exposure
separately and then averaged the visibilities obtained from two exposures. 
The errors in the calibrated visibilities of V3 were computed from
the errors given by the pipeline and the variations in the transfer
function calculated from three data sets of HD3689. 
The spectroscopic calibration was carried out as described in
Appendix~\ref{spec_calib}.

\section{Results}
\label{sect_res}

\begin{figure}
\begin{center}
\resizebox{\hsize}{!}{\rotatebox{0}{\includegraphics{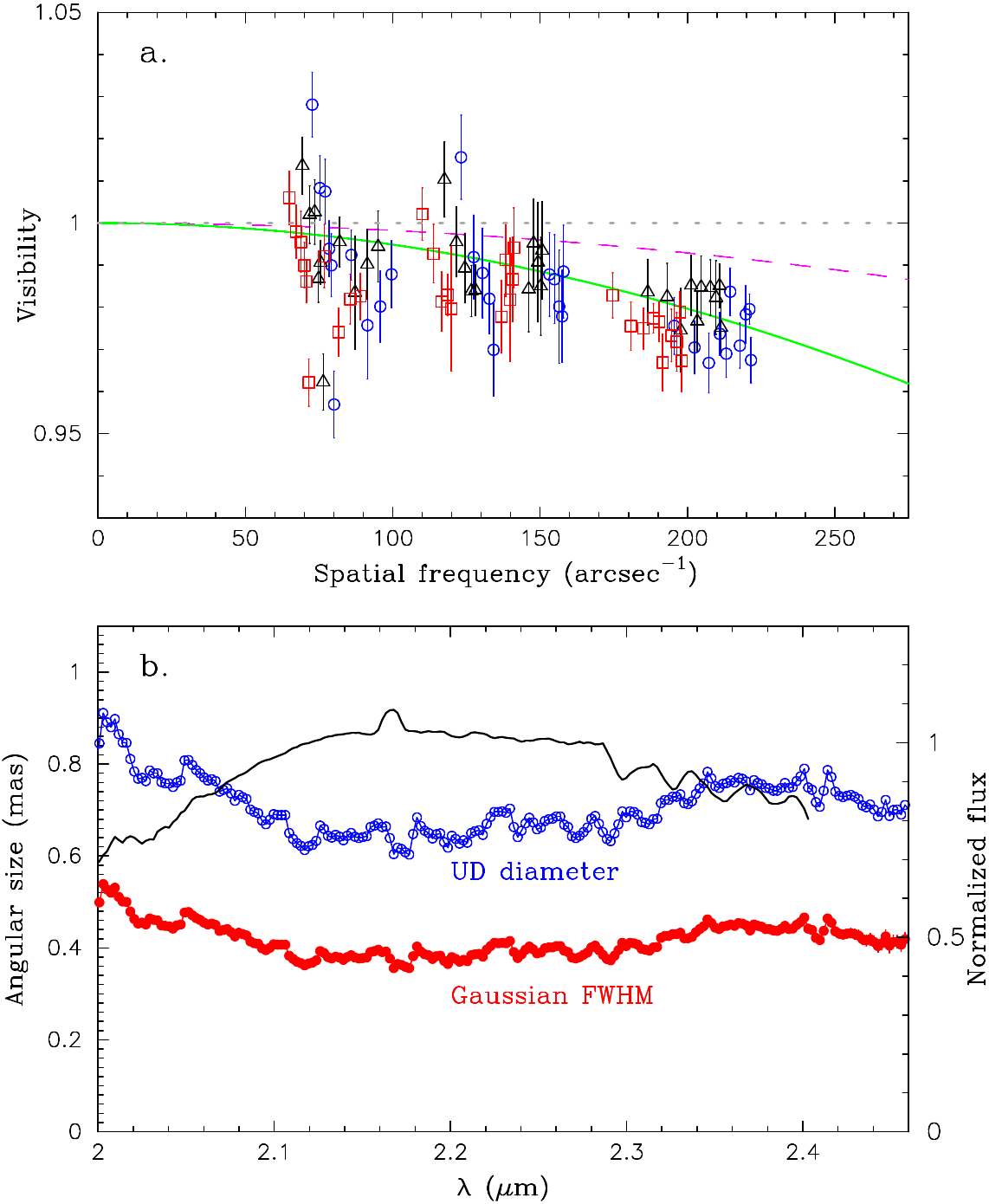}}}
\end{center}
\caption{
  GRAVITY observations of the AGB star V3 in the metal-poor globular cluster
  47~Tuc.
  {\bf a:} Visibilities observed at 2.1, 2.2, and 2.35~\mbox{$\mu$m}\ are
  plotted with the blue circles, black triangles, and red squares,
  respectively.
    The green solid line represents the visibility from the UD fit with an
    angular diameter of 0.65~mas derived at 2.2~\mbox{$\mu$m}, while the purple
    dashed line represents the uniform-disk visibility expected from the
    central star with a UD diameter of 0.38~mas. 
    The gray horizontal dotted line marks the visibility 1.
  {\bf b:}
    Wavelength dependence of 
    the FWHM and angular diameter obtained by the Gaussian fit
    and UD fit to the observed
    visibilities are shown by the red filled circles and blue open circles,
    respectively.
    The black solid line shows the
    normalized spectrum derived from the GRAVITY data. The peak seen at
  $\sim$2.165~\mbox{$\mu$m}\
  is not real, but caused instead by the Br $\gamma$ absorption line seen in the
  calibrator's spectrum, which was not entirely removed by our spectroscopic
  calibration.
  The spectrum is cut off at $\sim$2.4~\mbox{$\mu$m} because the spectrum of
  the proxy star used in the spectroscopic calibration  only extends to
  2.42~\mbox{$\mu$m}. 
}
\label{obsres}
\end{figure}

Figure~\ref{obsres}a shows the observed visibilities of V3 as a function
of spatial frequency. The visibilities obtained at the longest baselines
of 85-96~m (spatial frequencies of 180--220~arcsec$^{-1}$) are $\sim$0.97,
lower than 1 expected for a point source.
The data points at a spatial frequency of $\sim$150~arcsec$^{-1}$ are also
systematically slightly below 1, suggesting that the object is 
marginally resolved at a baseline length of $\sim$70~m.
This is the first spatially resolved observation of an individual 
star in a globular cluster. 
The differential phases and closure phases are zero within measurement
errors of $\sim$1\degr.
At shorter wavelengths of 1.59-1.76~$\mu$m, 
Hron et al. (\cite{hron15}) observed V3 with VLTI/PIONIER, 
but the object remained unresolved with an upper limit
of the UD diameter of 0.44~mas.

We fitted the visibilities observed at each wavelength with a Gaussian
  and a uniform disk (UD).
The resulting FWHM and UD diameter, shown in Fig.~\ref{obsres}b,
are $\sim$0.4~mas and $\sim$0.65~mas, respectively, between 2.1 and 
2.3~\mbox{$\mu$m}. 
The UD diameter of the central star is estimated to be 0.38~mas
from a stellar photospheric radius of 185~\mbox{$R_{\sun}$}
(Sect.~\ref{sect_model}) and the distance of 4.5~kpc.
The visibilities expected from the uniform disks with the diameters of 0.65
and 0.38~mas are also shown in Fig.~\ref{obsres}b. 
The measured UD diameter of $\sim$0.65~mas is 
1.7 times larger than that of the central star.
This may be interpreted as evidence of an extended dust envelope.

  However, we suggest that a different interpretation is also plausible. 
  The linear radius derived from the luminosity and effective temperature
  (see Sect.~\ref{sect_model}) 
  corresponds to the photospheric radius. 
  The apparent diameter of Mira stars can be larger than
  the photospheric diameter due to their extended atmospheres.
  Woodruff et al. (\cite{woodruff08}, \cite{woodruff09}) and
  Wittkowski et al. (\cite{wittkowski08}) showed that the angular diameter
  of Mira stars reaches a minimum at 1.2--1.4~\mbox{$\mu$m}, which approximately 
  corresponds to the photospheric diameter, while the uniform-disk diameter
  at 2.2~\mbox{$\mu$m}\ can be larger than that at 1.2--1.4~\mbox{$\mu$m}\ by a factor of
  up to $\sim$1.4.
  The UD diameter of 0.65~mas measured at 2.2~\mbox{$\mu$m}\ is 1.7 times
  larger than the photospheric size of 0.38~mas -- greater than
  the aforementioned factor of 1.4. However, if we adopt \mbox{$T_{\rm eff}$} = 2900~K
  instead of 3200~K
  in the calculation of the photospheric radius (Sect.~\ref{sect_model}),
  the stellar angular diameter is 0.46~mas and 
  the measured 2.2~\mbox{$\mu$m}\ UD diameter is 1.4 times larger than the
  photospheric size, as expected from the aforementioned studies. 
  This means that the observed visibilities can also be interpreted as due to
  the extended atmosphere of V3 with none or just a slight contribution from a
  dust shell at 2--2.45~\mbox{$\mu$m}.  
  It is possible that the data are explained by the contributions of 
  both the dust envelope and the extended atmosphere. 

The obtained FWHM and UD diameter 
tend to increase shortward of 2.1~\mbox{$\mu$m}\ and longward of
$\sim$2.3~\mbox{$\mu$m}. 
This can be explained as follows. The flux contribution of
the central star is lower at $\la$2.1~\mbox{$\mu$m} and
$\ga$2.3~\mbox{$\mu$m} because of the \mbox{H$_2$O} and CO absorption bands, 
as seen in the spectrum plotted in Fig.~\ref{obsres}b.
Therefore, 
if the observed data are interpreted as due to a dust envelope, 
its flux contribution 
is higher at $\la$2.1~\mbox{$\mu$m}\ and $\ga$2.3~\mbox{$\mu$m}, 
resulting in a larger apparent size. 
In addition, the extended atmosphere makes the 
central star appear to be larger in the CO and \mbox{H$_2$O}\ bands
than at $\sim$2.2~\mbox{$\mu$m}. This also leads to the observed increase in the
angular size at $\la$2.1 and $\ga$2.3~\mbox{$\mu$m}. 

\section{Dust shell modeling of the SED and GRAVITY visibilities}
\label{sect_model}

We collected (spectro)photometric data from the visible to the mid-infrared
for modeling of the SED and GRAVITY visibilities:
GAIA spectrum (Gaia Collaboration et al. \cite{gaia16}, \cite{gaia23}), 
2MASS $JHK_{s}$ photometry (Cutri et al. \cite{cutri03}),
Spitzer/IRS spectrum at 7--21~\mbox{$\mu$m}\
(Lebzelter et al. \cite{lebzelter06}\footnote{Downloaded from the Spitzer Data
Archive\\ https://irsa.ipac.caltech.edu/applications/Spitzer/SHA/}),
WISE photometry (Cutri et al. \cite{cutri14}), and AKARI photometry
(Ishihara et al. \cite{ishihara10}).
The data were corrected for the interstellar extinction with
$E(B-V) = 0.04$ (Harris \cite{harris96}) and $R_V = A_V/E(B-V) = 3.1$ 
using the wavelength dependence from Cardelli et al. (\cite{cardelli89}).
We obtained a bolometric flux of $5.1 \times 10^{-12}$~W~m$^{-2}$ by 
integrating the de-reddened SED, which corresponds to a bolometric luminosity
of 3200~\mbox{$L_{\sun}$}\ with the adopted distance of 4.5~kpc. 
Combined with an effective temperature (\mbox{$T_{\rm eff}$}) of 3200~K
adopted for the modeling below,
it results in a photospheric radius (\mbox{$R_{\rm ph}$}) of 185~\mbox{$R_{\sun}$}. 
While 47~Tuc V3 is a variable star as mentioned in
Sect.~\ref{sect_obs}, the collected (spectro)photometric data were 
taken at widely different epochs. 
Lebzelter \& Wood (\cite{lebzelter05a}) measured a $K$-band variability
amplitude of $\Delta K \approx 1$~mag, which affects the bolometric flux
and thus the luminosity. Lebzelter et al. (\cite{lebzelter14}) obtained
\mbox{$T_{\rm eff}$}\ = 3590~K and \mbox{$L_{\star}$} = 4590~\mbox{$L_{\sun}$}
for variability phase 0.18.
We checked the effects of the variations in \mbox{$T_{\rm eff}$} and
\mbox{$L_{\star}$}\ on the dust shell modeling, as described below.

We carried out radiative transfer modeling with
our Monte Carlo radiative transfer code mcsim\_mpi 
(Ohnaka et al. \cite{ohnaka06}).
For the parameters of the central star and grain properties, we adopted
those determined from the SED modeling of V3 by Mc11a. 
Those authors derived a value of \mbox{$T_{\rm eff}$}\ = 3153~K and a surface
gravity of $\mbox{$\log \varg$} = -0.25$. 
The radiation from the central star was approximated with the synthetic
spectrum of the MARCS model (Gustafsson et al. \cite{gustafsson08}\footnote{https://marcs.astro.uu.se/})
with parameters that are as close as possible to the above 
values of \mbox{$T_{\rm eff}$}\ and \mbox{$\log \varg$}. 
We selected the
MARCS model with \mbox{$T_{\rm eff}$}\ = 3200~K, \mbox{$\log \varg$}\ = 0.0,
\mbox{$M_{\star}$}\ = 1.0~\mbox{$M_{\sun}$}, \mbox{$\varv_{\rm micro}$}
(micro-turbulent velocity) = 2.0~\mbox{km s$^{-1}$}, along with the moderately
CN-cycled chemical composition with [Fe/H] = $-0.5$. The synthetic spectrum
was spectrally binned to match the spectral resolution of our GRAVITY data.

The SED modeling of metal-poor dusty stars in 47~Tuc reported by Mc11a 
shows that the primary dust
component is metallic iron. In the case of V3,
they concluded that the metallic iron fraction is 100\% based on the absence
of the spectral features due to silicate and corundum in the Spitzer/IRS
spectrum (Lebzelter et al. \cite{lebzelter06}) and the ground-based
spectrum (van Loon et al. \cite{vanloon06}). 
Therefore, we assumed metallic iron grains
in our modeling and the opacity was calculated with the complex
refractive index of Ordal et al. (\cite{ordal88}). The grain size
was assumed to be the standard Mathis, Rumpl, \& Nordsieck (MRN) 
distribution (Mathis et al. \cite{mathis77}),
with a minimum and maximum grain size of
0.005 and 0.25~\mbox{$\mu$m}, respectively.
We adopted the power-law radial density profile, described as
$\rho \propto r^{-p}$. 

The free parameters of the dust shell are the optical depth at
0.55~\mbox{$\mu$m} (\mbox{$\tau_{0.55}$}), the temperature
at the inner boundary radius (\mbox{$T_{\rm in}$}) of the dust shell,
and the power-law exponent, $p,$ of the density profile. 
The outer radius of the dust shell was fixed to be 5000 times larger
than the inner boundary radius, but the outer radius does not
affect the results of our modeling.
%
As described in Sect.~\ref{sect_res}, the extended atmosphere of V3 can 
make the central star's apparent size larger than the photospheric angular
diameter of 0.38~mas. To take this effect into account, the central star's
angular diameter was increased by a factor of $f$ in the calculation of the
model visibility.
We treated $f$ as a free parameter, adopting $f$ = 1 (no extended atmosphere),
1.35, and 1.7 (with the atmosphere's angular diameter equal to the observed
2.2~\mbox{$\mu$m}\ UD diameter of 0.65~mas). 

\begin{figure*}
\begin{center}
\resizebox{\hsize}{!}{\rotatebox{0}{\includegraphics{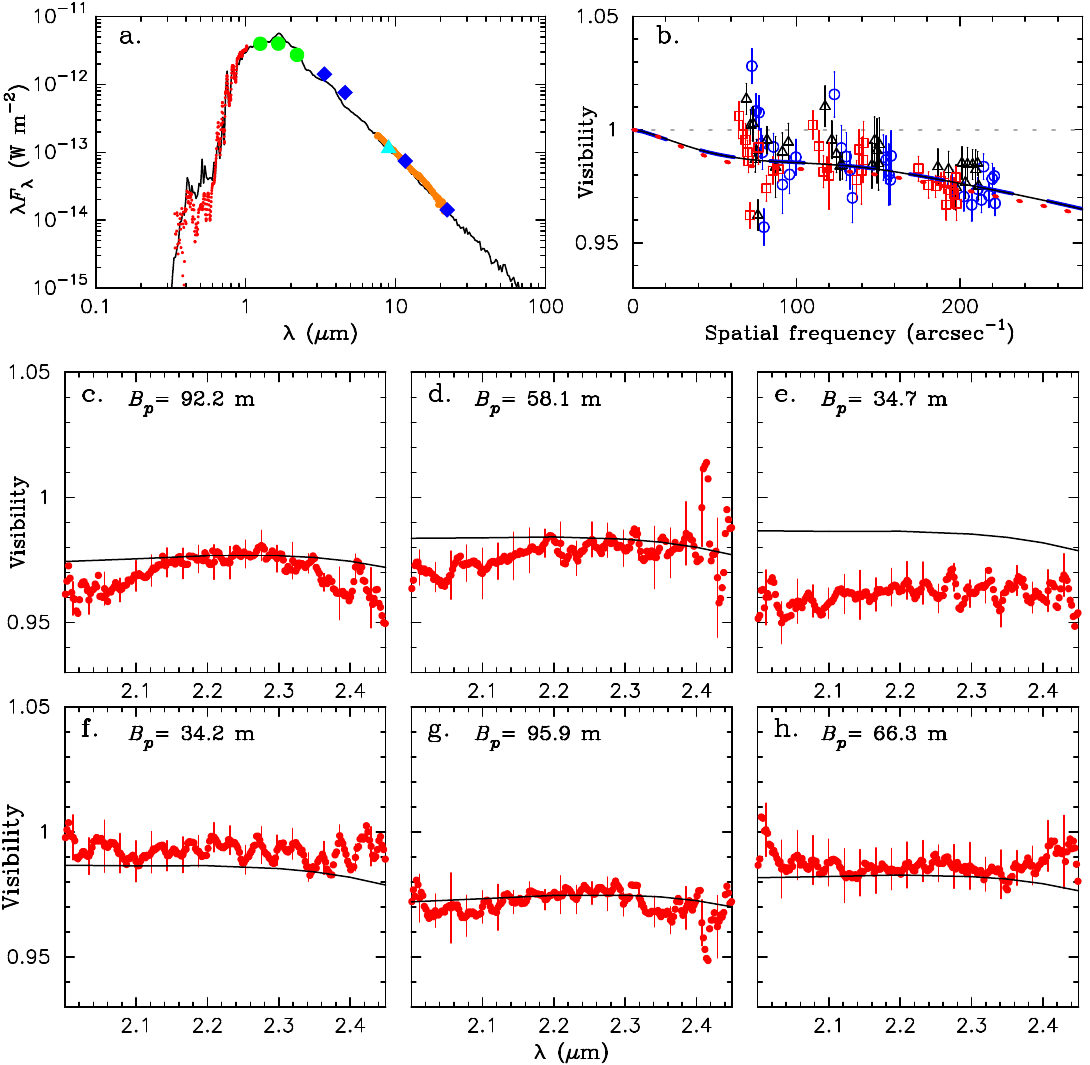}}}
\end{center}
\caption{
  Dust shell modeling of the SED and GRAVITY visibilities of 47~Tuc V3.
  {\bf a:} Comparison of the observed and model SEDs. 
  The solid line represents the best-fiting model with \mbox{$\tau_{0.55}$}\ =
    0.15
    and an inner boundary temperature of
      700~K. 
  The red dots, green circles,
  blue diamonds, light blue triangle, and thick orange  line
  represent the GAIA
  spectrum, 2MASS $JHK_{\rm s}$ photometric data, WISE photometric data,
  AKARI data, and Spitzer/IRS spectrum, respectively. 
  {\bf b:} Comparison of the observed and model visibilities
  as a function of spatial frequency.
  The blue circles, black triangles, and red squares
  correspond to the visibilities measured at
  2.1, 2.2, and 2.35~\mbox{$\mu$m}, respectively. 
  The blue long-dashed line, black solid line (almost entirely overlapping
  with the long-dashed line),
  and red dotted line
  represent the model visibilities predicted
  at 2.1, 2.2, and 2.35~\mbox{$\mu$m}, respectively. 
  The gray horizontal
    dotted line
  marks the visibility 1. 
  {\bf c}--{\bf h:} Comparison of the visibilities as a function of
  wavelength. In each panel, the red dots represent
  the observed data, while the black line represents the model. 
}
\label{v3model}
\end{figure*}

  Figure~\ref{v3model} shows a comparison of the observed SED and visibilities
  with those predicted by one of the best-fitting models, with 
  {$\tau_{0.55}$} = 0.15, $p$ = 2, \mbox{$T_{\rm in}$}\ = 700~K (corresponding to an
  inner radius of 35~\mbox{$R_{\rm ph}$}), and $f$ = 1.35. 
  The model visibilities plotted in Fig.~\ref{v3model}b indicate that 
  the visibility dips at spatial frequencies of $\la$60~arcsec$^{-1}$
  are attributed to the dust shell (i.e., it is resolved out at the
  baselines of our observations), 
 while the visibilities at higher spatial
  frequencies are attributed to the extended atmosphere of the central
  star. 
The observed SED is well
reproduced and the GRAVITY visibilities are reasonably fitted by the model,
given the errors in the data.
  The model predicts the decrease in the visibilities at $\la$2.1 and
  $\ga$2.3~\mbox{$\mu$m}\ to be less pronounced than the observed data
  as seen in Figs.~\ref{v3model}c, \ref{v3model}d, and \ref{v3model}g.
  This is because the wavelength dependence of the extended atmosphere
  is not included in our modeling. The slight dip at $\sim$2.3~\mbox{$\mu$m}\
  is due to the increase in the flux contribution of the dust shell, which
  makes the object's overall size slightly larger (i.e., visibility lower). 

Figure~\ref{v3model2} shows an alternative model, where the observed
visibilities are mostly explained by the extended atmosphere with only a
small contribution from the dust shell. 
This model is characterized by \mbox{$\tau_{0.55}$}\ = 0.1, $p$ = 2, \mbox{$T_{\rm in}$}\ = 500~K,
and $f$ = 1.7. It should be noted that the inner boundary temperature
is very low (corresponding to an inner radius of 90~\mbox{$R_{\rm ph}$}) and, therefore, 
the flux contribution of the dust shell at 2--2.45~\mbox{$\mu$m}\ is $\la$0.5\%.
The figure shows that the fit to the observed data is comparable to
the model shown in Fig.~\ref{v3model}. 
We found that the dust shell models with \mbox{$\tau_{0.55}$}\ = 0.1--0.25, \mbox{$T_{\rm in}$}\ =
500--800~K, and $p$ = 2--2.5 are able to aptly reproduce the data. Models
with lower \mbox{$T_{\rm in}$}\ have higher values of $f$, as, for instance,
seen in the above two models. 

If we adopt the higher \mbox{$T_{\rm eff}$} and \mbox{$L_{\star}$} determined
by Lebzelter et al. (\cite{lebzelter14}), the fit to the GAIA data in the
visible is very poor, because the 3500~K MARCS model predicts the 
TiO absorption to be much less shallow than observed.
If we only fit the SED longward of
$\sim$1~\mbox{$\mu$m}\ and the GRAVITY visibilities,
the best-fitting models have 
\mbox{$\tau_{0.55}$}\ = 0.05--0.15 and \mbox{$T_{\rm in}$}\ = 500--800~K with $p$ = 2--2.5. 
While the dust shell
is slightly more optically thin, its inner boundary temperature
is not significantly affected by the uncertainties in
\mbox{$T_{\rm eff}$}\ and \mbox{$L_{\star}$}. 

  Our modeling shows that 
  the GRAVITY visibilities of V3 can be explained by the
  dust shell and/or the extended atmosphere of the central star.
  While the atmospheric extension, $f,$ remains ambiguous between 1 and 1.7,
  the inner boundary temperature ranges from 500 to 800~K.
  These temperatures are noticeably lower
than the condensation temperature of $\sim$1500~K of corundum
(\mbox{Al$_2$O$_3$}) and Fe-free silicates
such as forsterite (\mbox{Mg$_2$SiO$_4$}) and enstatite (\mbox{MgSiO$_3$})
found in nearby oxygen-rich AGB stars (Khouri et al. \cite{khouri16};
Ohnaka et al. \cite{ohnaka16}, \cite{ohnaka17};
Ohnaka \& Adam \cite{adam19}).
Mc11a obtained an inner boundary temperature
of 1000~K for V3 from their SED fitting
(with 100\% metallic iron as mentioned above) --
lower than 1500~K but still higher than 500--800~K.
%
  The inner boundary temperatures of 500--800~K correspond to radii of
  90--23~\mbox{$R_{\rm ph}$},
which is much larger than $\sim$2 stellar radii reported
in the aforementioned studies of the nearby oxygen-rich AGB stars. 
The mid-infrared spectrum
of V3 does not show a trace of the silicates or corundum 
in spite of their higher condensation temperatures compared to
metallic iron. 
Mc11a pointed out that it is not clear how
metallic iron condenses before the Mg-rich (and Fe-poor) silicates or
corundum. This question becomes more serious, with the inner boundary
temperature shown to be even lower based on our modeling of the GRAVITY data. 

We calculated the ratio of the acceleration due to the radiation pressure
on metallic iron grains to the gravity using:
\[
\beta_{\rm dust} = 1146 \frac{L_{\star}}{10^4 \mbox{$L_{\sun}$}} \frac{1}{M_{\star} (\mbox{$M_{\sun}$})} 
\frac{Q}{0.2} \frac{1}{a (\mbox{$\mu$m})}
\frac{1}{\rho_{\rm bulk} ({\rm g \, cm^{-3})}}
\frac{1}{r_{\rm gd}}, 
\]
where $L_{\star}$, $M_{\star}$, $Q$, $a$, $\rho_{\rm bulk}$, and
$r_{\rm gd}$ are 
the luminosity of the central star, its mass, the flux-mean opacity, 
the grain radius, the bulk density of dust grains, and the gas-to-dust
ratio, respectively (Yamamura et al. \cite{yamamura00}). 
We calculated the flux-mean opacity of metallic iron dust with the stellar
spectrum used in our modeling for different grain sizes of 0.005, 0.1, and
0.25~\mbox{$\mu$m}. Assuming a gas-to-dust ratio of 1000 (Mc11a), 
$\rho_{\rm bulk}$ = 7~g~cm$^{-3}$, and a stellar mass of
0.6--0.7~\mbox{$M_{\sun}$}\ (McDonald et al. \cite{mcdonald11b};
Lebzelter et al. \cite{lebzelter14}), 
the ratio is 0.55, 2.5, and 3.7 for the grain size of 0.005, 0.1,
and 0.25~\mbox{$\mu$m}, respectively, with a ratio of 1 reached for a grain size of
$\sim$0.05~\mbox{$\mu$m}.
This means that grains smaller than $\sim$0.05~\mbox{$\mu$m}\
are not blown away by the radiation pressure.

On the other hand, Mc11a suggested that
metallic iron grains in the stars in 47~Tuc are small or elongated and/or
they condense more efficiently than at the solar metallicity, to account for the mass-loss rate and the wind terminal velocity. 
As an example,
they presented a case with iron grains five times smaller than the standard
MRN size distribution, which means that the maximum grain size is
0.05~\mbox{$\mu$m}. However, they noted that both effects -- the smaller iron grain
size and elongated grains -- are relevant and, therefore, they could not 
specify the size of the small grains; it seems likely that the size 
is in the aforementioned range,
where the radiation pressure on iron grains does not lead to mass loss. 
This implies, as Mc11a concluded, that
the mass loss in V3 as well as other cool evolved stars in 47~Tuc may not be
driven by radiation pressure on dust grains, but by some other mechanism(s) instead. 
In this case, dust formation is a mere result -- and not the
source -- of the mass loss.

\section{Conclusion}
\label{sect_concl}

Our VLTI/GRAVITY observations have spatially resolved, for the first time,
an individual star, V3, in the metal-poor globular cluster 47~Tuc
at 2--2.45~\mbox{$\mu$m}.
  The observed GRAVITY data 
translate into a Gaussian FWHM angular size of $\sim$0.4~mas
and a UD diameter of $\sim$0.7~mas.
Our radiative transfer modeling suggests that
the observed SED and GRAVITY data can be explained by an optically thin dust
shell with \mbox{$\tau_{0.55}$}\ = 0.05--0.25, consisting of metallic iron
grains, likely with effects of the extended atmosphere of the central star. 
The dust condensation temperature was found to be 500--800~K, 
significantly lower than found in nearby Galactic AGB stars. 

GRAVITY observations at shorter and longer baselines than those seen in
the present data are necessary to 
better constrain the properties of the dust shell and the
extended atmosphere of the central star, respectively. 
To further study the formation of metallic iron grains,
thermal-infrared interferometric
observations will be useful. The VLTI/MATISSE instrument allows us to
spatially resolve V3 simultaneously at 3-4.2~\mbox{$\mu$m}, 4.5-5~\mbox{$\mu$m}, and
8-13~\mbox{$\mu$m}. In particular, observations in the 10~\mbox{$\mu$m}\
region are important for examining whether there is a trace of
grain species other than metallic iron in the spatially resolved data.

\begin{acknowledgement}
We thank the ESO Paranal team for supporting our VLTI observations.
K.O. acknowledges the support of the Agencia Nacional de 
Investigaci\'on Cient\'ifica y Desarrollo (ANID) through
the FONDECYT Regular grant 1210652. 
This research made use of the \mbox{SIMBAD} database, 
operated at the CDS, Strasbourg, France. 
This publication makes use of data products from the Two Micron All Sky
Survey, which is a joint project of the University of Massachusetts and the
Infrared Processing and Analysis Center/California Institute of Technology,
funded by the National Aeronautics and Space Administration and the National
Science Foundation.
This publication makes use of data products from the Wide-field Infrared
Survey Explorer, which is a joint project of the University of California, Los
Angeles, and the Jet Propulsion Laboratory/California Institute of Technology,
funded by the National Aeronautics and Space Administration.
This research has made use of the NASA/IPAC Infrared Science Archive, which is
funded by the National Aeronautics and Space Administration and operated by
the California Institute of Technology.
This research is based on observations with AKARI, a JAXA project with the
participation of ESA.This work has made use of data from the European Space
Agency (ESA) mission
{\it Gaia} (\url{https://www.cosmos.esa.int/gaia}), processed by the {\it Gaia}
Data Processing and Analysis Consortium (DPAC,
\url{https://www.cosmos.esa.int/web/gaia/dpac/consortium}).
Funding for the DPAC
has been provided by national institutions, in particular the institutions
participating in the {\it Gaia} Multilateral Agreement.

\end{acknowledgement}

\clearpage
\appendix

\section{Observation log}
\label{appendix_obslog}

Table~\ref{obslog} shows the summary of our GRAVITY observations of
47~Tuc V3 and the calibrator HD3689. 

\begin{table}
\caption {
Summary of our VLTI/GRAVITY observations of 47~Tuc V3.
}

\begin{tabular}{l c c c c r}\hline
\# & $t_{\rm obs}$ & $B_{\rm p}$ & PA     & Seeing   & $\tau_0$ \\ 
   & UTC         &  (m)       & (\degr) & (\arcsec) &  (ms)  \\
\hline
\multicolumn{6}{c}{2023 October 14}\\
\multicolumn{6}{c}{AT configuration: D0-G2-K0-J3}\\
\hline
\multicolumn{6}{c}{47~Tuc V3}\\
\multicolumn{6}{c}{DIT = 10~s, N$_{\rm f}$ = 32, N$_{\rm exp}$ = 2}\\
\hline
1 & 00:29:47 & 92.2/58.1/34.7/ & 13/15/$-72$/ & 0.65 & 6.6 \\
  &         & 34.3/95.9/66.3  & 10/34/47     &      &     \\
2 & 01:16:17 & 91.3/57.5/37.2/ & 22/25/$-63$/ & 0.44 & 10.5 \\
  &          & 33.9/95.6/67.0  & 19/45/58     &      &     \\
3 & 02:05:59 & {89.7}/56.5/39.6/  & 32/35/$-53$/ & 0.62 & 5.6 \\
  &          & 33.4/95.1/67.7  & 28/57/70     &      &     \\
4 & 02:52:47 & 87.6/55.2/{41.4}/ & 42/44/$-44$/ & 0.55 & 4.4 \\
  &          & 32.6/94.2/68.1  & 37/68/82     &      &     \\
5 & 03:44:14 & 84.6/53.3/43.1/ & 52/55/$-35$/ & 0.44 & 5.3 \\
  &          & 31.5/92.8/68.3  & 47/80/$-86$  &      &     \\
\hline
\multicolumn{6}{c}{Calibrator HD3689}\\
\multicolumn{6}{c}{DIT = 10~s, N$_{\rm f}$ = 32, N$_{\rm exp}$ = 2}\\
\hline
C1 & 00:04:50 & 91.0/57.4/32.1/ & 6/8/97/ & 0.65 & 7.2 \\
   &          & 33.7/95.9/66.1  & 2/25/37 &      &     \\
C2 & 00:53:08 & 90.4/57.0/34.8/ & 15/18/$-72$/ & 0.57 & 7.5 \\
   &          & 33.5/95.7/66.8  & 12/37/49     &      &     \\
C3 & 01:40:08 & 89.4/56.4/37.1/ & 25/27/$-63$/ & 0.53 & 12.0 \\
   &          & 33.1/95.3/67.4  & 21/48/61     &      &      \\
\hline
\label{obslog}
\vspace*{-5mm}

\end{tabular}
\tablefoot{
$B_{\rm p}$: Projected baseline length.  PA: Position angle of the baseline 
vector projected onto the sky. 
DIT: Detector Integration Time.  $N_{\rm f}$: Number of frames in each 
exposure.  $N_{\rm exp}$: Number of exposures. 
The seeing and the coherence time ($\tau_0$) were measured in the visible band. 
}
\end{table}

\section{Spectroscopic calibration of the GRAVITY spectrum}
\label{spec_calib}

The spectroscopic calibration to remove the telluric lines and instrumental
effects was carried out as:
\[
F_{\rm sci}^{\rm true} = F_{\rm sci}^{\rm obs} \times F_{\rm cal}^{\rm true}/F_{\rm cal}^{\rm obs},
\]
where $F_{\rm sci (cal)}^{\rm true}$ and $F_{\rm sci (cal)}^{\rm obs}$
denote the true and observed spectra of the science target (sci) or 
the calibrator (cal) HD3689, respectively. 
To approximate the true spectrum of HD3689 (F6V), we used the spectrum of
$\xi$~Peg taken with the InfraRed Telescope Facility (IRTF Spectral
Library\footnote{http://irtfweb.ifa.hawaii.edu/\~{}spex/IRTF\_Spectral\_Library/}, Rayner et al. \cite{rayner09}) because its spectral type and
luminosity class F6V is the same as that of HD3689.
The IRTF spectrum taken with a spectral resolution of 
$\lambda/\Delta \lambda = 2000$ was convolved to match the spectral
resolution of 200 of our GRAVITY data and then used for 
$F_{\rm cal}^{\rm true}$ in the above spectroscopic calibration. 
The calibrated spectrum of 47~Tuc V3 was normalized with the flux at
2.2~\mbox{$\mu$m}, where the flux is the least affected by the \mbox{H$_2$O}\ and CO bands. 

\section{Alternative model with little contribution of the dust shell}
\label{alt_model}

\begin{figure*}
\begin{center}
\resizebox{\hsize}{!}{\rotatebox{0}{\includegraphics{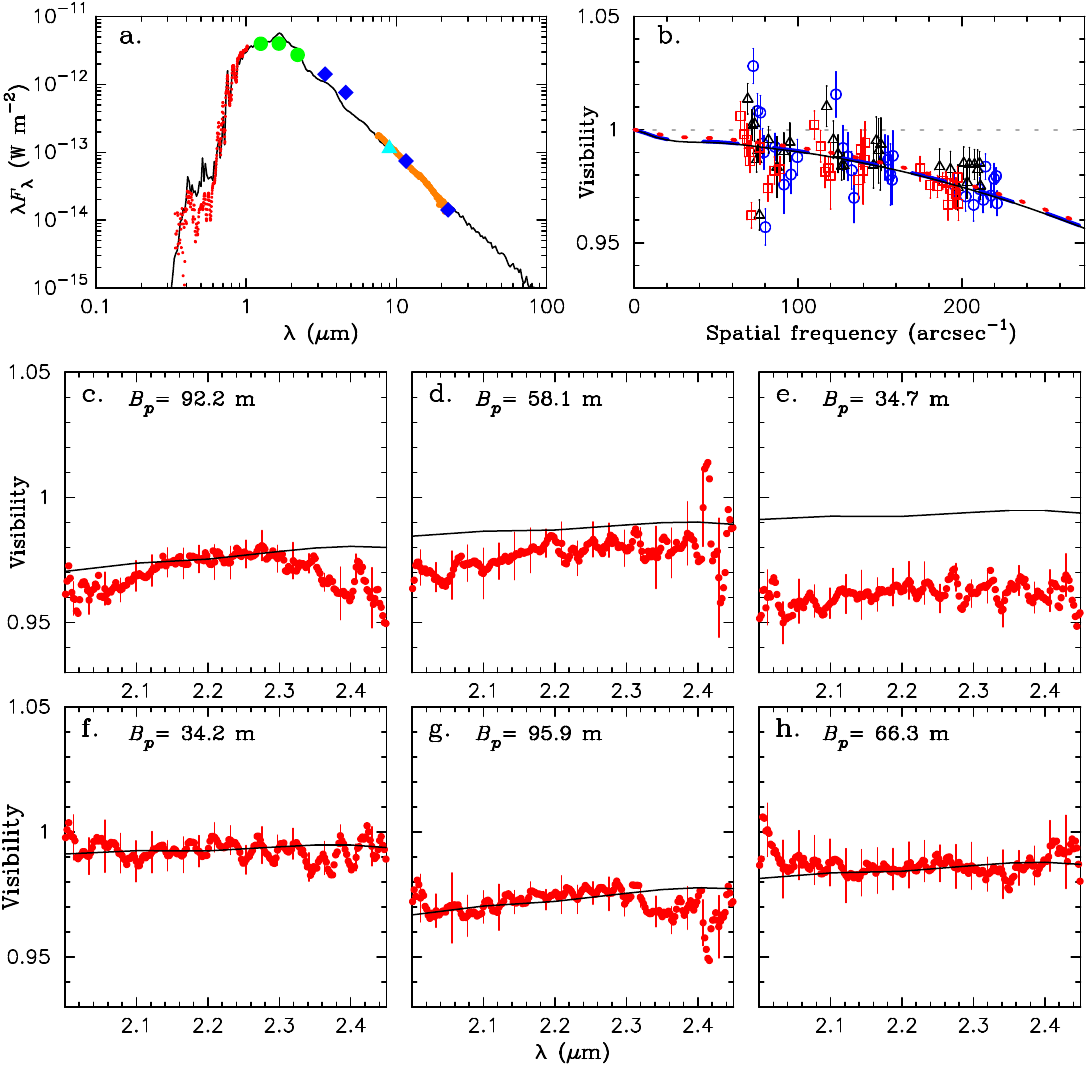}}}
\end{center}
\caption{
  Alternative model for the SED and GRAVITY visibilities of 47~Tuc V3 
  with little flux contribution from the dust shell, shown in the same
  manner as in Fig.~\ref{v3model}.
}
\label{v3model2}
\end{figure*}

Figure~\ref{v3model2} shows a comparison of the observed SED and GRAVITY
visibilities with a model where the flux contribution of the dust shell
at 2--2.45~\mbox{$\mu$m}\ is smaller than $\sim$0.5\%. In this case,
the visibilities observed at 2--2.45~\mbox{$\mu$m}\ are accounted for by
the atmosphere of the central star extending to 1.7~\mbox{$R_{\rm ph}$}. 

\end{document}